\documentclass[12pt,a4paper,dvips]{article}
\usepackage{a4p}
\usepackage{cite,mcite}
\usepackage{graphicx}
\usepackage{physics }
\usepackage{l3_title,ifthen}
\usepackage{rotating}
\journalname{Phys. Lett. B}

\usepackage{Lep}
\Lep{2}

\preprint{99-119}
\date{August 26, 1999}

\newlength{\capindent}
\newlength{\capwidth}
\newlength{\figwidth}
\newcommand{\icaption}[2][!*!,!]{\hspace*{\capindent}%
  \begin{minipage}{\capwidth}
    \ifthenelse{\equal{#1}{!*!,!}}%
      {\caption{#2}}%
      {\caption[#1]{#2}}
  \end{minipage}}

\newcommand{\WW}{{\rm WW}}

\newcommand{\qqll}{{\rm q \bar q\ell^+\ell^- }}

\newcommand{\llnn}{{\rm \ell^+\ell^- \nu\bar\nu}}
\newcommand{\llll}{{\rm \ell^+\ell^-\ell^{\prime +}\ell^{\prime -} }} 
\newcommand{\qqnn}{{\rm q\bar q\nu\bar\nu}}
\newcommand{\qqqq}{{\rm q\bar qq^\prime\bar{q}^\prime}}

\begin{document}
\setlength{\unitlength}{1mm}
\begin{titlepage}
\title{Study of Z Boson Pair Production 
 in \boldmath{\epem} Collisions at LEP at \boldmath{$\sqrt{s}=189\GeV\,$}}
\author{The L3 Collaboration}
\begin{abstract}
The pair production of Z bosons is studied using the data
collected by the L3 detector at LEP in 1998 in $\epem$ collisions at
a centre--of--mass energy of $189\GeV$. All the visible final states are
considered and the cross section of this process is measured to be
$0.74^{+0.15}_{-0.14}\,{(\rm stat.)}\,{\pm
  0.04}\,{(\rm syst.)}\,\mathrm{pb}$.
Final states
containing b quarks are enhanced by a dedicated selection
and their  production cross section is found to be
$0.18^{+0.09}_{-0.07}\,{(\rm stat.)}\,\pm 0.02\,{\rm
  (syst.)}\,\mathrm{pb}$. 
Both results are in agreement with the Standard
Model predictions. 
Limits on anomalous couplings between neutral gauge bosons are derived
from these measurements.
\end{abstract}
\submitted

\end{titlepage}

\section{Introduction}                                         

Since 1997 LEP is running at  centre--of--mass energies, $\sqrt{s}$,
above the production threshold of Z boson pairs. This process is
of particular interest 
as it constitutes an irreducible background for the search of the
Standard Model Higgs boson and to several other processes predicted by
theories beyond the Standard Model.  In addition it allows the
investigation of possible triple neutral gauge boson couplings, ZZZ and
ZZ$\gamma$~\cite{hagiwara}, forbidden at tree level in the Standard 
Model. 

The experimental investigation of ZZ production is made difficult by
its rather low cross section, compared with competing processes
that constitute large and sometimes irreducible
backgrounds.  
The existence at threshold of this process was established at 
$\sqrt{s} = 183\GeV$~\cite{zzl3183}.
In the following, the analysis of the data collected at $189\GeV$
is described. The measurement of the cross section is presented together
with that of final states containing b quarks. Limits on anomalous
couplings among neutral gauge bosons are derived.

\section{Data and Monte Carlo Samples}

The data were collected in 1998 by the L3
detector~\cite{l3_00} at
$\mathrm{\sqrt{s}}=188.7\GeV$, and amount to an integrated luminosity of 
176 pb$^{-1}$.
This energy will be denoted as $189\GeV$ in the following.

The EXCALIBUR~\cite{exca} Monte Carlo is used to
generate events belonging to both the signal and the background
neutral--current four--fermion processes. Background from 
fermion--pair production is described making use of
PYTHIA\,5.72~\cite{pythia} 
($\rm e^+ e^- \rightarrow q \overline q  (\gamma)$), 
KORALZ\,4.02~\cite{koralz} ($\rm e^+ e^- \rightarrow \mu^+ \mu^-  (\gamma)$
and $\rm e^+ e^- \rightarrow \tau^+ \tau^-  (\gamma)$ )
and BHWIDE~\cite{bhwide} ($\rm e^+ e^-\rightarrow e^+ e^-  (\gamma)$).
Background from charged--current four--fermion processes
is generated with EXCALIBUR for $\rm e\nu_\e q\bar q'$ and
$\rm \ell^+\nu_\ell \ell^-\bar \nu_\ell$ with $\rm \ell=e,\mu,\tau$ and
KORALW\,1.21~\cite{koralw} for  $\WW$ production.
Contributions from multiperipheral processes
are modelled by
PHOJET\,1.05c~\cite{phojet} ($\rm e^+ e^- \rightarrow e^+ e^- q \overline q$)
and DIAG36\,\cite{diag36} ($\rm e^+ e^- \rightarrow e^+ e^- \ell^+\ell^- $).

The L3 detector response is simulated using the GEANT 3.15
program~\cite{geant}, which takes into account the effects of energy loss,
multiple scattering and showering in the detector. 
Time dependent detector inefficiencies, as measured in the data taking
period, are reproduced in these simulations.

The definition of the Z pair signal is unchanged with respect to the
generator level 
phase--space cuts of the $183\GeV$ analysis~\cite{zzl3183}. 
Those requirements are summarised as follows: the invariant mass of both
the generated fermion pairs must be between $70\GeV$ and
$105\GeV$. This criterion has to be satisfied by at least one of the two
possible pairings of four same flavour fermions. In the case in which
fermion pairs can originate from a charged--current process 
($\rm u\bar u d \bar d $, $\rm c\bar c s \bar s $   
and $\rm\nu_\ell\bar{\nu_\ell} \ell^+\ell^-$, with $\rm \ell=e,\mu,\tau$) 
the masses of the fermion pairs which could 
come from W decays  are required to be either below  $75\GeV$
or above $85\GeV$. Events with electrons in the final state are
rejected if $|\cos{\theta_{\rm e}}| > 0.95$, where $\theta_{\rm e}$
is the electron polar angle.

The expected cross sections for the different final states are
computed  with EXCALIBUR.
A total
cross section of 0.662 pb is expected.
 In this calculation  $\alpha_s=0.119$~\cite{pdg} 
is included for the QCD vertex corrections.
The cross section for states with at least one b--quark pair amounts to
0.178\,pb.

\section{Event Selection}

All the visible final states of the Z pair decay are investigated,
with criteria similar to those used at $183\GeV$~\cite{zzl3183}. 
All selections are based on the identification of two fermion pairs
each with a mass close to the Z boson mass.
The selections are
modified to take into account the different background composition at
the higher $\sqrt{s}$ and the changed signal topology due to the
larger boost of the Z bosons. This boost leads to acollinear and acoplanar
fermion pairs.

\subsection{\boldmath{$\qqll$} Channel}

A dedicated selection is performed for each of the final states
$\rm{q\bar{q}e^+e^-}$, $\rm{q\bar{q}\mu^+\mu^-}$ and
$\rm{q\bar{q}\tau^+\tau^-}$ after the application of a 
common preselection. This requires at least five charged tracks,
15 calorimetric clusters and a visible energy of more than
0.4$\sqrt{s}$ together with two same flavour identified leptons.

Electrons are identified from energy depositions in the electromagnetic
calorimeter whose shower shape is compatible with those initiated by
an electron or a photon. At least one electron should have a matched
track.
Muons are reconstructed from tracks in the muon spectrometer pointing
to the interaction vertex.  
Energy depositions in the calorimeters consistent with a minimum ionising
particle (MIP) which have an associated track are also accepted as
second muon candidates.
For the $\rm{q\bar{q}\tau^+\tau^-}$ channel both a particle--based  and a
jet--based selection are performed. In the
first, tau leptons are identified via their 
decay into isolated electrons or muons, or as an isolated
low--multiplicity jet with  one or three tracks and unit charge. In
the jet--based selection, the event is forced into four--jets using the
DURHAM~\cite{durham} algorithm. Two of the jets must each have less
than four tracks.  These jets are considered as tau candidates, but at
least one of them  must coincide  with a tau candidate defined in the
particle--based selection. 

In the electron and the muon channels both the lepton and the jet pair must
have an opening angle of at least $120^\circ$, tightened to
$130^\circ$ for the  taus.
The invariant mass of
the jet--jet  and the lepton--lepton system after performing a kinematic
fit, which  imposes energy  and momentum conservation, must be within
$70\GeV$ and $120\GeV$. The events are then subject 
to the DURHAM algorithm requiring  $\rm{lnY_{34}}$ to be greater than $-6.0$
for the electron and tau channels and  $-6.5$ for the muon one.
$\rm Y_{34}$ is the jet resolution parameter for which the event
is changed from a four--jet to a  three--jet
topology. Furthermore, the  visible energy in the electron channel
must be at least $0.8\sqrt{s}$ and between  0.6$\sqrt{s}$ and
0.9$\sqrt{s}$ for the jet--based tau selection.

Additional requirements are applied in the tau selection to reduce
the radiative $\rm{q\bar{q}(\gamma)}$ background rejecting events
containing a photon of energy larger than $30\GeV$. 
Semileptonic WW events are rejected by requiring 
the transverse missing momentum to be lower than
$40\GeV$ in events with no identified
electron or muon with energy larger than $40\GeV$.

The kinematic fit is repeated on events that pass at least one of the
four selections described above with the extra constraint of equal
invariant masses for the jet--jet and lepton--lepton systems. 
 The distribution of the invariant mass arising
from the fit,
$ M_{5C}$, is shown in Figure~1.
Table~1 summarises the yield of  this selection.

\begin{table}[ht]
  \begin{center}
    \begin{tabular}{|c|r|r|r|c|}
      \hline
      Selection & Data & Signal MC & Background MC &
      Efficiency \\
      \hline   
      $\qqll$ & 15  & $10.9\pm 0.2$ & $   5.4\pm 0.2$ & $61\%$ \\  
      $\qqnn$ & 40  & $15.7\pm 0.5$ & $  18.9\pm 0.8$ & $49\%$ \\
      $\llnn$ & 3   & $0.7 \pm 0.1$ & $   1.2\pm 0.1$ & $22\%$ \\ 
      $\llll$ & 2   & $0.7 \pm 0.0$ & $   0.6\pm 0.1$ & $37\%$ \\
      $\qqqq$ & 163 & $24.2\pm 0.4$ & $ 153.0\pm 1.2$ & $43\%$ \\
      \hline
    \end{tabular}
    \caption{Data, signal and background Monte Carlo events selected
      by each analysis and their efficiency. The $\qqnn$ entries are reported for a 
      selection requirement of 0.5 on the neural network output. The
      $\llnn$ figures refer only to electrons and muons. The Monte
      Carlo statistical uncertainties are given on signal and
      background expectations.} 
  \end{center}
\end{table}

\subsection{\boldmath{$\qqnn$} Channel}

High multiplicity hadronic events with more than three charged
tracks and at least 15 calorimetric energy clusters are selected.
The event invariant mass must exceed $50 \GeV$. These cuts reduce
contributions from purely leptonic two--fermion final states, as well
as two--photon interactions, while keeping a significant fraction of
hadronic events from $\rm q \bar{q} (\gamma)$ and W--pair
production. These latter contributions are further reduced by
requiring the visible mass to be less than $130 \GeV$ and the mass
recoiling against the hadronic system to exceed $50 \GeV$.

In addition,
the transverse momentum is required to be greater than $5\GeV$ and
the longitudinal momentum to be smaller than 40\% of the visible energy.
The energy deposition in the forward calorimeters must not
exceed $10\GeV$ and the missing momentum vector must be at least
$16^\circ$ away from the beam axis. No electrons, muons
or photons with energies above $20\GeV$ are allowed in the
event and the energy in a 25$^\circ$ azimuthal sector
around the missing energy direction, $E_{25}$, is required to be smaller than
$30 \GeV$.

A total of 299 events satisfy the selection criteria with 23 and 266
events expected from the signal and background Monte Carlo simulations
respectively. The dominating background is due to charged--current
four--fermion processes. To differentiate further the ZZ signal from
the remaining background a neural network is constructed.
The inputs to the neural network include event shape variables which
help to distinguish two--jet from three--jet topologies, the
sum of invariant and missing masses, the masses of the two jets, the
total missing momentum and $E_{25}$. The use of the neural network
increases the signal fraction in the selected sample to approximately
60\% for large neural network output values, as demonstrated in Figure~2a.
The efficiency and the yield of this selection are reported in Table~1
for a cut at 0.5, while the full spectrum is used for the cross section
determination.

\subsection{\boldmath{$\llnn$} Channel}

The selection for $\llnn$ is optimised for electron and muon pairs
identified as in Section 3.1 and characterised by an
invariant mass, $M_{\ell\ell}$, between $85\GeV$ and $95\GeV$. The
requirement on the associated track for electrons is dropped and
MIPs are not considered.

In the electron channel only events with a visible energy between
$75\GeV$ and $98\GeV$ are selected; this requirement is loosened to the
range $65\GeV -140\GeV$ for muons. The opening angle of the
two electrons must be below 166$^\circ$ and from $143^\circ$ to
$172^\circ$ for the two muons. In order to reduce the
background from radiative Bhabha scattering and purely leptonic decays of W
pairs, the recoil mass to the electron pair is required to be
less than $95\GeV$.
The background from
other resonant and non--resonant four--fermion processes is reduced by performing
 a kinematic fit imposing the Z mass to the
visible pair of leptons and recalculating their
four--momenta. The recoil mass, $M_{rec}$, after the fit is
required to be less than $98\GeV$ for electrons and in
excess of  $84\GeV$ but not larger than $98\GeV$ for muons. The transverse momentum has
to lie in the range from $4\GeV$ to $29\GeV$.

The spectrum of the sum of $M_{\ell\ell}$ and $M_{rec}$, without
applying the kinematic fit, peaked around twice the Z mass is presented in
Figure~2b. Table~1 lists the efficiencies and the yield of the selection.
No contribution in the $\tau^+\tau^-\nu_{\tau}\bar{\nu}_{\tau}$
signal channel is expected.

\subsection{\boldmath{$\llll$} Channel}

This selection is based on events with at least four loosely identified
leptons of a minimum energy of  $3\GeV$  and the subsequent study of
just one pair of them.

First 
a low multiplicity event preselection is applied, requiring at least
two tracks but less than 15 calorimetric clusters, with a total 
visible energy between $0.2\sqrt{s}$ and $1.3\sqrt{s}$. 
Electrons and muons are identified as described in Section 3.1.
Low angle electromagnetic showers ($|\cos{\theta}| > 0.95$) without a matching track
are also accepted as electrons.
Tau candidates are identified as low multiplicity hadronic jets with
either one or three 
tracks in a cone of $10^\circ$ half opening angle.  To reject
hadronic jets, the energy between 
$10^\circ$ and $30^\circ$ around the tau direction must not exceed
half of the energy in a cone of $10^\circ$ half opening angle.
To increase the selection efficiency, MIPs are also accepted.

If there are more than four lepton candidates, the four most
consistent with energy and momentum conservation are chosen.
Events are then required to have at least one electron, 
muon, tau or muon--MIP pair. Low angle  electrons are not
considered in this procedure. If more than one such 
pair is possible, the one with the invariant mass, $M_{\ell\ell}$,
closest to the Z mass is chosen.

Both $M_{\ell\ell}$ and the recoil mass, $M_{rec}$, to this
selected lepton pair are 
required to lie between $70\GeV$ and $105\GeV$.
The data and Monte Carlo distributions for $M_{\ell\ell} + M_{rec}$
are shown in Figure~2c. Table~1 summarises the
total yield of the selection.

\subsection{\boldmath{$\qqqq$} Channel}

The four--jet
selection has to cope with the large QCD and W pair production
backgrounds. 
High multiplicity hadronic events are selected by requiring a
visible energy between 0.6$\sqrt{s}$ and  
1.4$\sqrt{s}$ together with parallel and perpendicular imbalances below 0.3$\sqrt{s}$.
Events with an identified electron, muon or photon with energy in excess
of $65\GeV$ are discarded.

The events forced to four jets with the DURHAM algorithm are 
subjected to a constrained fit which rescales the jets to balance
momentum while imposing energy conservation.
This fit reduces greatly the dependence on the calorimeter energy scale.
A first neural network~\cite{ww} is then applied to distinguish events
 with four genuine quark jets from those  with two quark jets and two
jets from gluon radiation.  
A cut on the output of this
neural network rejects QCD background selecting a hadronic sample
enhanced in W and Z pairs.

A second network uses five variables 
to  distinguish Z pairs from W pairs by means of their mass difference
after the dijet pairings are chosen to minimise the dijet mass
difference. The variables are
the reconstructed dijet mass,  the maximum 
and minimum energy in any jet, the average number of charged tracks
per jet and the dijet mass difference. 
A large portion of four--jet ZZ decays contains at least one b quark 
pair, which  provides significant distinguishing power from W pair decays. A
sixth variable, a
b--tag discriminant~\cite{h183}, is added to the network.
The output of this network  is shown in Figure~2d. A cut on the
background enhanced region below 0.2 is applied and the rise
above 0.8 is due to the b--tagged events. The performances of the
analysis are summarised in Table~1.

A simpler sequential analysis is also performed for this channel, taking
advantage of the different boost of  Z and W pairs and hence
investigating  just the two dijet opening angles, the dijet mass
difference and the dijet mean mass. Results 
compatible with the previous approach are obtained limited by a
lower purity  due to the absence of b--tagging.

\section{Measurement of the ZZ Cross Section}

A binned maximum
likelihood fit to each of the variables displayed in Figures~1 and~2
allows the determination of the ZZ cross section in the individual
final states, as listed in  Table~2. These are in agreement with the
Standard Model values reported in the same table. Assuming these
predictions as the relative weights of different channels, the 
ZZ cross section,  $\sigma_{\rm ZZ}$, is found to be:
\begin{displaymath}
\sigma_{\rm ZZ}=0.74^{+0.15}_{-0.14}\,{\rm
  pb},
\end{displaymath}
in good agreement with the expected cross section within the signal
definition cuts of   
0.662\,pb. The uncertainties are only statistical.

In the calculation of the cross section the effect of the cross talk
between the separate channels is found to be negligible.

\begin{table}[ht]
\begin{center}
\begin{tabular}{|c|c|c|c|c|c|}
\hline
\rule{0pt}{12pt}  & $\qqll$ & $\qqnn$ & $\llnn$ & $\llll$ & $\qqqq$ \\
\hline
\rule{0pt}{12pt} Measured cross section (pb)&
$0.096^{+ 0.039}_{-0.033}$ &
$0.23^{+ 0.07}_{-0.06}$ &
$0.054^{+ 0.059}_{-0.040}$ &
$0.017^{+ 0.025}_{-0.015}$ &
$0.32^{+ 0.14}_{-0.13}$ \\
\rule{0pt}{12pt} Expected cross section (pb)&
0.102 &
0.179 &
0.027 &
0.011 &
0.316 \\
\hline
\end{tabular}
\caption{Result of the individual cross section fits.}
\end{center}
\end{table}

\section{Study of Systematic Errors}

The systematic uncertainties are grouped in correlated
and uncorrelated sources among the channels.
The correlated sources of systematic errors are the background 
cross sections, 
the LEP energy and the energy scale of 
the detector.  As they modify the shapes of the investigated
distributions, their effect is evaluated
performing a new fit to calculate $\sigma_{\rm ZZ}$ once their
values are scaled to the extremes  listed in Table~3.  
An  uncertainty  of 2\% is attributed to the measured
cross section to take into account the difference of the assumed relative
weights of the different channels with respect to those
obtained with the GRC4F~\cite{grace} Monte Carlo  generator and to
parametrise the other uncertainties related  to their calculation.

Four sources of systematic uncertainty are uncorrelated among the
channels and modify the shapes of some of the discriminating
distributions. 
The jet resolution
and the charged track multiplicity for the $\qqqq$ selection are
scaled as in Table~3.
The b--tag procedure and the Monte Carlo description 
of b--hadron jets  are taken into account by reweighting the value
of the b--tag discriminant.
Finally
the  Bhabha Monte Carlo  is not
sufficient to estimate  the background to the  $\llnn$
channel. This estimate  is obtained from
a fit to the shapes of some selection variables whose uncertainty
contributes to the total systematic uncertainty.
Two additional sources of systematic uncertainty are propagated to the total cross
section: the 
Monte Carlo statistics and lepton identification. Their uncertainties
listed in Table~4  do not affect the shape of the
discriminating distributions. 

The individual and combined systematic errors are listed in Table~3.

The measured cross section is then:
\begin{displaymath}
\sigma_{\rm ZZ}=0.74^{+0.15}_{-0.14}\,{(\rm stat.)}\,{\pm
  0.04}\,{(\rm syst.)}\,\mathrm{pb}.
\end{displaymath}

\begin{table}[ht]
  \begin{center}
    \begin{tabular}{|c|c|r|r|}
      \hline
      Systematic Source & Variation & $\delta\sigma_{\rm ZZ}$ (pb) &
      $\delta\sigma_{\rm ZZ\ra b\bar{b}X}$   (pb)  \\
      \hline
      \rule{0pt}{12pt}Correlated sources                       &          &                  &   \\
      \hline
      \rule{0pt}{12pt}Lep energy                    &$40\,\MeV$&$ < 0.01$&$ < 0.01$ \\
      \rule{0pt}{12pt}WW cross section              & 2\%      &$   0.01$&$ < 0.01$ \\
      \rule{0pt}{12pt}Four--jet rate                 & 5\%      &$   0.01$&$   0.01$ \\
      \rule{0pt}{12pt}W$\rm{e}\nu$ cross section    & 10\%     &$   0.01$&$ < 0.01$ \\
      \rule{0pt}{12pt}Four--fermion cross section   & 5\%      &$ < 0.01$&$   0.01$ \\
      \rule{0pt}{12pt}Energy scale                  & 2\%      &$   0.01$&$   0.01$ \\
      \rule{0pt}{12pt}Theory predictions            & 2\%      &$   0.01$&$ < 0.01$ \\
      \hline                                                                 
      \rule{0pt}{12pt}Uncorrelated sources                       &      &              &   \\
      \hline                                                                 
      \rule{0pt}{12pt}Jet resolution  ($\qqqq$)     & 2\%      &$   0.01$&$ < 0.01$ \\
      \rule{0pt}{12pt}Charge multiplicity ($\qqqq$) & 1\%      &$ < 0.01$&$ < 0.01$ \\
      \rule{0pt}{12pt}B--tag  ($\qqqq$)             & see text &$   0.01$&$   0.01$ \\
      \rule{0pt}{12pt}Bhabha background ($\llnn$)   & see text &$   0.01$&$     - $\\
      \rule{0pt}{12pt}Monte Carlo statistics        & see text &$   0.02$&$   0.01$ \\
      \rule{0pt}{12pt}Lepton identification         & see text &$   0.01$&$   0.01$ \\
      \hline                                                                 
      \rule{0pt}{12pt}Total                         &          &$   0.04$&$   0.02$ \\
      \hline
    \end{tabular}
    \caption{Systematic uncertainties on  $\sigma_{\rm ZZ}$ and 
      $\sigma_{\rm ZZ\ra b\bar{b}X}$.}
  \end{center}
\end{table}

\begin{table}[ht]
  \begin{center}
    \begin{tabular}{|c|c|c|c|}
      \hline
      Channel & Systematic Source     & Background uncertainty & Signal uncertainty \\
      \hline
      $\qqll$ & Monte Carlo statistics& 4.7\% & 1.7\% \\
              & Lepton identification & 2.7\% & 4.1\% \\
      \hline
      $\qqnn$ & Monte Carlo statistics& 4.0\% & 3.0\% \\
      \hline
      $\llnn$ & Monte Carlo statistics& 8.0\% & 8.0\% \\
              & Lepton identification & 5.0\% & 4.0\% \\
      \hline
      $\llll$ & Monte Carlo statistics& 21.4\% & 2.1\% \\
              & Lepton identification & 10.1\% & 4.3\% \\
      \hline
      $\qqqq$ & Monte Carlo statistics& 0.8\% & 1.7\% \\
      \hline
    \end{tabular}
    \caption{Sources of uncorrelated  systematic uncertainties.}
  \end{center}
\end{table}

In terms of the NC02 cross section in which only the two
conversion diagrams are considered for the  double--resonant Z
pair production, the cross section reads
\begin{displaymath}
\sigma^{\rm NC02}_{\rm ZZ}=0.73^{+0.15}_{-0.14}\,{(\rm stat.)}\,{\pm
  0.04}\,{(\rm syst.)}\,\mathrm{pb},
\end{displaymath}
to be compared with a Standard Model expectation of 0.65\,pb
calculated with the YFSZZ~\cite{YFSZZ} package.

\section{b Quark Content in ZZ Events}

It is of particular interest to investigate the rate 
of ZZ events with  b
quark content. The production of the minimal or a supersymmetric Higgs
boson
would manifest via an enhancement of these events and their
study complements the dedicated search for such
processes~\cite{h183}. The expected Standard Model cross
section for ZZ$\rm \ra b\bar{b}X$  final states is  0.178\,pb.  

The investigation of the ZZ$\rm \ra b\bar{b}X$ events proceeds by
complementing the analyses of the $\qqnn$ and $\qqll$ final states
described above with a further variable describing the b quark content in
the event~\cite{h183}. 
Three variables are then considered for each final state: 
${ M_{5C}}$ for the $\qqll$ and the neural network output for the
$\qqnn$ analysis together with  the b--tag evaluated
for each of the two hadronic jets.
The combination of each of the sets of these three variables into a
single discriminant proceeds as follows. First the variables are 
mapped to achieve uniform distributions for the background. Then the
product of their observed values is calculated event by event. Finally
the 
confidence level is calculated for the product of three uniformly distributed
quantities to be less than the observed product.
This confidence level is expected to be low for signal and flat
for background. The final discriminant is the negative logarithm
of this confidence level and is shown in Figure~3.
Figure~2d shows the ${\rm b\bar bq\bar{q}}$ response from the
neural network used to select the $\qqqq$ final states.

The cross section calculation is performed as above. Results for the
individual channels are listed in Table~5. The combined result for 
 $\sigma_{\rm ZZ\ra b\bar{b}X}$ is: 
\begin{displaymath}
\sigma_{\rm ZZ\ra b\bar{b}X}=0.18^{+0.09}_{-0.07}\,{\rm
  (stat.)}\,\pm 0.02\,{\rm (syst.)}\,\mathrm{pb}.
\end{displaymath}
This result agrees with the Standard Model expectation and differs
from zero at 99.9\% confidence level.
In the fit the other ZZ final states are fixed to their Standard Model
expectations. The systematic uncertainties are evaluated in the same way as for
the total cross section and are presented in Table~3.
Figure~4 displays the measured total and ${\rm b\bar{b}X}$ cross
sections and their expected 
evolution with $\sqrt{s}$.

\begin{table}[ht]
\begin{center}
\begin{tabular}{|c|c|c|c|}
\hline
\rule{0pt}{12pt}  & $\rm b\bar{b}\ell^+\ell^-$ & $\rm b\bar{b}\nu\bar{\nu}$ &
$\rm q \bar q b\bar{b}$ \\
\hline
\rule{0pt}{12pt} Measured cross section (pb)&
$0.020^{+ 0.022}_{-0.014}$ &
$0.044^{+ 0.046}_{-0.036}$ &
$0.111^{+ 0.076}_{-0.062}$ \\
\rule{0pt}{12pt} Expected cross section (pb)&
0.021 &
0.039 &
0.118 \\
\hline
\end{tabular}
\caption{Result of the individual  ZZ$\rm \ra b\bar{b}X$ cross section fits.}
\end{center}
\end{table}

\section{Anomalous Couplings}

A parametrisation of the  ZZZ and ZZ$\gamma$ anomalous couplings is
given in Reference~\cite{hagiwara}. 
Assuming on-shell production of a pair of Z bosons, only four couplings
$f_i^{\mathrm V}~(i=4,5 ; {\mathrm V}=\gamma,\Zo)$, 
where the V superscript corresponds to an anomalous coupling $\rm ZZV$, 
may be different from zero. At tree level these couplings are zero in
the Standard Model. They are independent from the $h_i^\Zo$
couplings that parametrise the possible anomalous ZZ$\gamma$ vertex~\cite{hagiwara},
probed by the $\epem\ra\Zo\gamma$ process~\cite{zgamma}.

In order to calculate the impact of anomalous couplings
on the measured distributions in 
the process
$\rm e^+e^- \rightarrow f \bar{f} f' \bar{f}'$, 
the EXCALIBUR generator is extended~\cite{madridpaper}.
The matrix elements of the Standard Model are supplemented 
by an additional term  containing anomalous couplings, 
${\cal M}_{\mathrm{AC}}(\{p^{\nu}\},\lambda,f_i^{\mathrm V})$ \cite{hagiwara}, 
where $ \{p^{\nu}\}$  
represents the phase space variables and $\lambda$ the helicities of 
initial and final state fermions.
Four--fermion Monte Carlo distributions for non--zero
anomalous couplings are then obtained by
reweighting
each event with the factor 
\begin{displaymath}
  W(\{p^{\nu}\},f_i^{\mathrm V})  \equiv 
 \frac { \frac{1}{4} {\displaystyle \sum_{\lambda}}{ \left | {\displaystyle
         ( {\cal M}_{\mathrm{4f}}(\{p^{\nu}\},\lambda) + {\cal
 M}_{\mathrm{AC}}(\{p^{\nu}\},\lambda,f_i^{\mathrm V}) )~}
                                                     \right |^2} }
       { \frac{1}{4} {\displaystyle \sum_{\lambda}}{ \left | {\displaystyle 
           {\cal M}_{\mathrm{4f}}(\{p^{\nu}\},\lambda)~} \right |^2}},
\end{displaymath}
where $ {\cal M}_{\mathrm{4f}}(\{p^{\nu}\},\lambda)$ is the Standard Model amplitude for the 
four--fermion final states. An
average over initial state and a sum over final state helicities are
carried out.
Initial state radiation is taken into account by evaluating 
the event weight at the centre--of--mass of the four--fermion system.
Figure~1 displays the effects of an
anomalous value of $f_4^{\gamma}$ obtained by reweighting with this
technique the four--fermion Monte Carlo events selected by the $\qqll$ analysis. 

The effects of anomalous couplings not only change
the ZZ  cross section  but also the shape of the
distributions. 
Using the distributions given in
Figures~1 and 2, a binned
maximum likelihood fit is therefore performed
for each of the anomalous couplings $f_i^{\mathrm V}$, fixing the
others to zero. 

The results of all these fits are
compatible with the Standard Model and 95\% confidence level limits on the
couplings are set as follows:
\begin{displaymath}
-1.9  \leq f_4^{\Zo}    \leq 1.9;\   \ 
-5.0  \leq f_5^{\Zo}    \leq 4.5;\   \
-1.1  \leq f_4^{\gamma} \leq 1.2;\   \ 
-3.0  \leq f_5^{\gamma} \leq 2.9.
\end{displaymath}
These limits are still valid for off--shell ZZ production where
additional couplings are possible. 
The small asymmetries in these limits are due to the interference
term between the anomalous coupling diagram and the Standard 
Model diagrams.
Systematic uncertainties on signal and background cross sections are
taken into account in the derivation of the limits.

These limits  improve by nearly a factor two previously published results at
$\sqrt{s}=183\GeV$~\cite{zzl3183},  which are included in the present analysis.

%
%
\section*{Acknowledgements}

We thank the CERN accelerator divisions for the
excellent performance and the continuous and successful upgrade of the
LEP machine.  
We acknowledge the contributions of the engineers  and technicians who
have participated in the construction and maintenance of this experiment.


\bibliographystyle{l3stylem}

%
%
\newpage
\typeout{   }     
\typeout{Using author list for the 189 ZZ paper only}
\typeout{Using author list for the 189 ZZ paper only}
\typeout{Using author list for the 189 ZZ paper only}
\typeout{Using author list for the 189 ZZ paper only}
\typeout{Using author list for the 189 ZZ paper only}
\typeout{Using author list for the 189 ZZ paper only}
\typeout{Using author list for the 189 ZZ paper only}
\typeout{$Modified: Mon Aug 23 15:15:13 1999 by clare $}
\typeout{!!!!  This should only be used with document option a4p!!!!}
\typeout{   }
%
%
%
%
%
%

\newcount\tutecount  \tutecount=0
\def\tutenum#1{\global\advance\tutecount by 1 \xdef#1{\the\tutecount}}
\def\tute#1{$^{#1}$}
\tutenum\aachen            
\tutenum\nikhef            
\tutenum\mich              
\tutenum\lapp              
\tutenum\basel             
\tutenum\lsu               
\tutenum\beijing           
\tutenum\berlin            
\tutenum\bologna           
\tutenum\tata              
\tutenum\ne                
\tutenum\bucharest         
\tutenum\budapest          
\tutenum\mit               
\tutenum\debrecen          
\tutenum\florence          
\tutenum\cern              
\tutenum\wl                
\tutenum\geneva            
\tutenum\hefei             
\tutenum\seft              
\tutenum\lausanne          
\tutenum\lecce             
\tutenum\lyon              
\tutenum\madrid            
\tutenum\milan             
\tutenum\moscow            
\tutenum\naples            
\tutenum\cyprus            
\tutenum\nymegen           
\tutenum\caltech           
\tutenum\perugia           
\tutenum\cmu               
\tutenum\prince            
\tutenum\rome              
\tutenum\peters            
\tutenum\salerno           
\tutenum\ucsd              
\tutenum\santiago          
\tutenum\sofia             
\tutenum\korea             
\tutenum\alabama           
\tutenum\utrecht           
\tutenum\purdue            
\tutenum\psinst            
\tutenum\zeuthen           
\tutenum\eth               
\tutenum\hamburg           
\tutenum\taiwan            
\tutenum\tsinghua          
{
\parskip=0pt
\noindent
{\bf The L3 Collaboration:}
\ifx\selectfont\undefined
 \baselineskip=10.8pt
 \baselineskip\baselinestretch\baselineskip
 \normalbaselineskip\baselineskip
 \ixpt
\else
 \fontsize{9}{10.8pt}\selectfont
\fi
\medskip
\tolerance=10000
\hbadness=5000
\raggedright
\hsize=162truemm\hoffset=0mm
\def\r{\rlap,}
\noindent

M.Acciarri\r\tute\milan\
P.Achard\r\tute\geneva\ 
O.Adriani\r\tute{\florence}\ 
M.Aguilar-Benitez\r\tute\madrid\ 
J.Alcaraz\r\tute\madrid\ 
G.Alemanni\r\tute\lausanne\
J.Allaby\r\tute\cern\
A.Aloisio\r\tute\naples\ 
M.G.Alviggi\r\tute\naples\
G.Ambrosi\r\tute\geneva\
% ...
\newpage
\begin{list}{A}{\itemsep=0pt plus 0pt minus 0pt\parsep=0pt plus 0pt minus 0pt
                \topsep=0pt plus 0pt minus 0pt}
\item[\aachen]
 I. Physikalisches Institut, RWTH, D-52056 Aachen, FRG$^{\S}$\\
 III. Physikalisches Institut, RWTH, D-52056 Aachen, FRG$^{\S}$
\item[\nikhef] National Institute for High Energy Physics, NIKHEF, 
     and University of Amsterdam, NL-1009 DB Amsterdam, The Netherlands
\item[\mich] University of Michigan, Ann Arbor, MI 48109, USA
\item[\lapp] Laboratoire d'Annecy-le-Vieux de Physique des Particules, 
     LAPP,IN2P3-CNRS, BP 110, F-74941 Annecy-le-Vieux CEDEX, France
\item[\basel] Institute of Physics, University of Basel, CH-4056 Basel,
     Switzerland
\item[\lsu] Louisiana State University, Baton Rouge, LA 70803, USA
\item[\beijing] Institute of High Energy Physics, IHEP, 
  100039 Beijing, China$^{\triangle}$ 
\item[\berlin] Humboldt University, D-10099 Berlin, FRG$^{\S}$
\item[\bologna] University of Bologna and INFN-Sezione di Bologna, 
     I-40126 Bologna, Italy
\item[\tata] Tata Institute of Fundamental Research, Bombay 400 005, India
\item[\ne] Northeastern University, Boston, MA 02115, USA
\item[\bucharest] Institute of Atomic Physics and University of Bucharest,
     R-76900 Bucharest, Romania
\item[\budapest] Central Research Institute for Physics of the 
     Hungarian Academy of Sciences, H-1525 Budapest 114, Hungary$^{\ddag}$
\item[\mit] Massachusetts Institute of Technology, Cambridge, MA 02139, USA
\item[\debrecen] Lajos Kossuth University-ATOMKI, H-4010 Debrecen, Hungary$^\P$
\item[\florence] INFN Sezione di Firenze and University of Florence, 
     I-50125 Florence, Italy
\item[\cern] European Laboratory for Particle Physics, CERN, 
     CH-1211 Geneva 23, Switzerland
\item[\wl] World Laboratory, FBLJA  Project, CH-1211 Geneva 23, Switzerland
\item[\geneva] University of Geneva, CH-1211 Geneva 4, Switzerland
\item[\hefei] Chinese University of Science and Technology, USTC,
      Hefei, Anhui 230 029, China$^{\triangle}$
\item[\seft] SEFT, Research Institute for High Energy Physics, P.O. Box 9,
      SF-00014 Helsinki, Finland
\item[\lausanne] University of Lausanne, CH-1015 Lausanne, Switzerland
\item[\lecce] INFN-Sezione di Lecce and Universit\'a Degli Studi di Lecce,
     I-73100 Lecce, Italy
\item[\lyon] Institut de Physique Nucl\'eaire de Lyon, 
     IN2P3-CNRS,Universit\'e Claude Bernard, 
     F-69622 Villeurbanne, France
\item[\madrid] Centro de Investigaciones Energ{\'e}ticas, 
     Medioambientales y Tecnolog{\'\i}cas, CIEMAT, E-28040 Madrid,
     Spain${\flat}$ 
\item[\milan] INFN-Sezione di Milano, I-20133 Milan, Italy
\item[\moscow] Institute of Theoretical and Experimental Physics, ITEP, 
     Moscow, Russia
\item[\naples] INFN-Sezione di Napoli and University of Naples, 
     I-80125 Naples, Italy
\item[\cyprus] Department of Natural Sciences, University of Cyprus,
     Nicosia, Cyprus
\item[\nymegen] University of Nijmegen and NIKHEF, 
     NL-6525 ED Nijmegen, The Netherlands
\item[\caltech] California Institute of Technology, Pasadena, CA 91125, USA
\item[\perugia] INFN-Sezione di Perugia and Universit\'a Degli 
     Studi di Perugia, I-06100 Perugia, Italy   
\item[\cmu] Carnegie Mellon University, Pittsburgh, PA 15213, USA
\item[\prince] Princeton University, Princeton, NJ 08544, USA
\item[\rome] INFN-Sezione di Roma and University of Rome, ``La Sapienza",
     I-00185 Rome, Italy
\item[\peters] Nuclear Physics Institute, St. Petersburg, Russia
\item[\salerno] University and INFN, Salerno, I-84100 Salerno, Italy
\item[\ucsd] University of California, San Diego, CA 92093, USA
\item[\santiago] Dept. de Fisica de Particulas Elementales, Univ. de Santiago,
     E-15706 Santiago de Compostela, Spain
\item[\sofia] Bulgarian Academy of Sciences, Central Lab.~of 
     Mechatronics and Instrumentation, BU-1113 Sofia, Bulgaria
\item[\korea] Center for High Energy Physics, Adv.~Inst.~of Sciences
     and Technology, 305-701 Taejon,~Republic~of~{Korea}
\item[\alabama] University of Alabama, Tuscaloosa, AL 35486, USA
\item[\utrecht] Utrecht University and NIKHEF, NL-3584 CB Utrecht, 
     The Netherlands
\item[\purdue] Purdue University, West Lafayette, IN 47907, USA
\item[\psinst] Paul Scherrer Institut, PSI, CH-5232 Villigen, Switzerland
\item[\zeuthen] DESY, D-15738 Zeuthen, 
     FRG
\item[\eth] Eidgen\"ossische Technische Hochschule, ETH Z\"urich,
     CH-8093 Z\"urich, Switzerland
\item[\hamburg] University of Hamburg, D-22761 Hamburg, FRG
\item[\taiwan] National Central University, Chung-Li, Taiwan, China
\item[\tsinghua] Department of Physics, National Tsing Hua University,
      Taiwan, China
\item[\S]  Supported by the German Bundesministerium 
        f\"ur Bildung, Wissenschaft, Forschung und Technologie
\item[\ddag] Supported by the Hungarian OTKA fund under contract
numbers T019181, F023259 and T024011.
\item[\P] Also supported by the Hungarian OTKA fund under contract
  numbers T22238 and T026178.
\item[$\flat$] Supported also by the Comisi\'on Interministerial de Ciencia y 
        Tecnolog{\'\i}a.
\item[$\sharp$] Also supported by CONICET and Universidad Nacional de La Plata,
        CC 67, 1900 La Plata, Argentina.
\item[$\diamondsuit$] Also supported by Panjab University, Chandigarh-160014, 
        India.
\item[$\triangle$] Supported by the National Natural Science
  Foundation of China.
\item[\dag] Deceased.
\end{list}
}
\vfill





\newpage

%
%

\newpage
\begin{figure}[p]
\begin{center}
\includegraphics[width=17.0truecm]{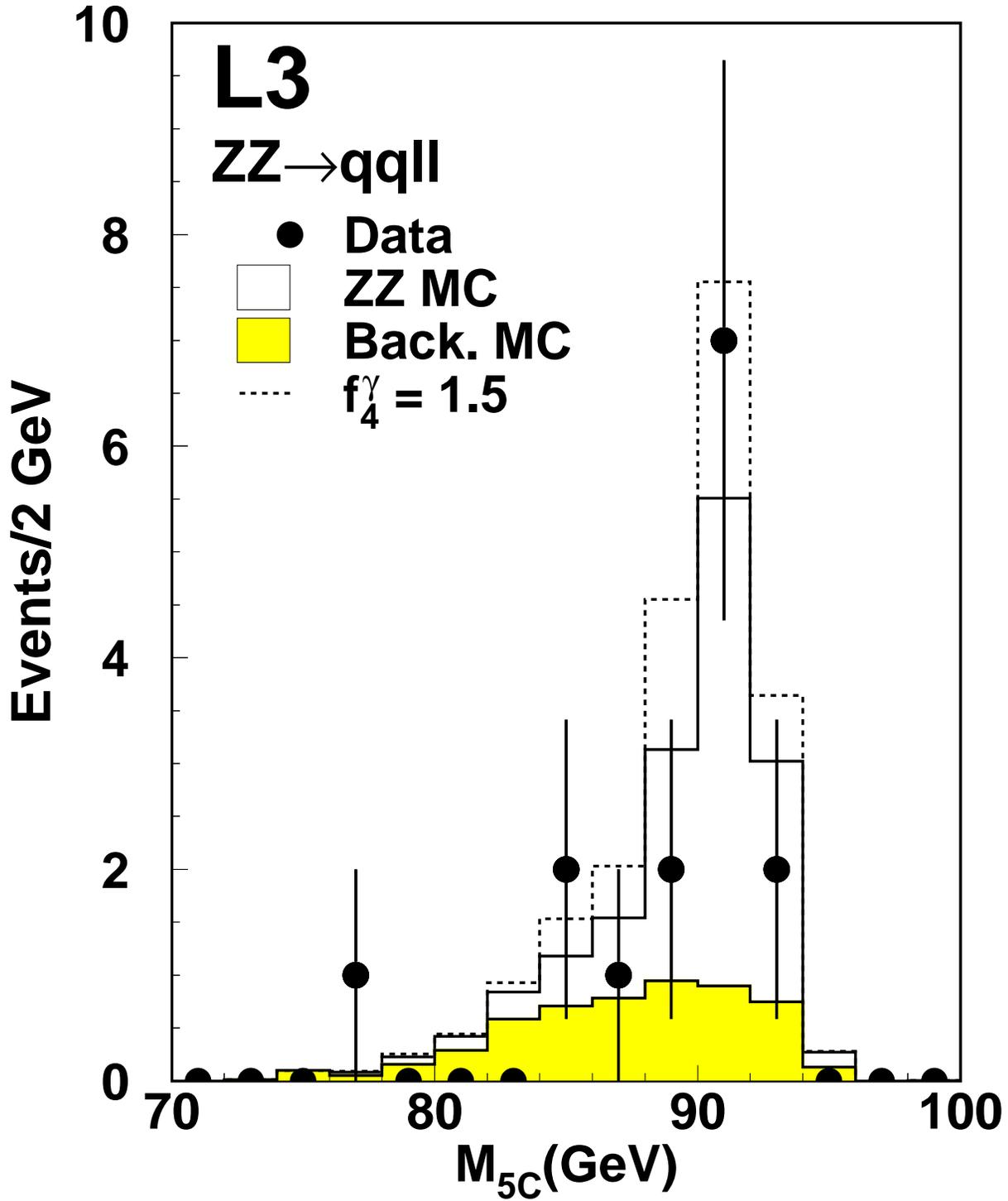}
\caption{Invariant mass after a kinematic fit, $M_{5C}$, of the lepton
  pair for the $\qqll$ selected 
  events. The effect of an anomalous ZZ$\gamma$ vertex is also shown
  for a value of its coupling $f_4^\gamma = 1.5$.}
\label{fig:1}
\end{center}
\end{figure}

\newpage
\begin{figure}[p]
\begin{center}
\begin{tabular}{cc}
\mbox{\includegraphics[width=8.5truecm]{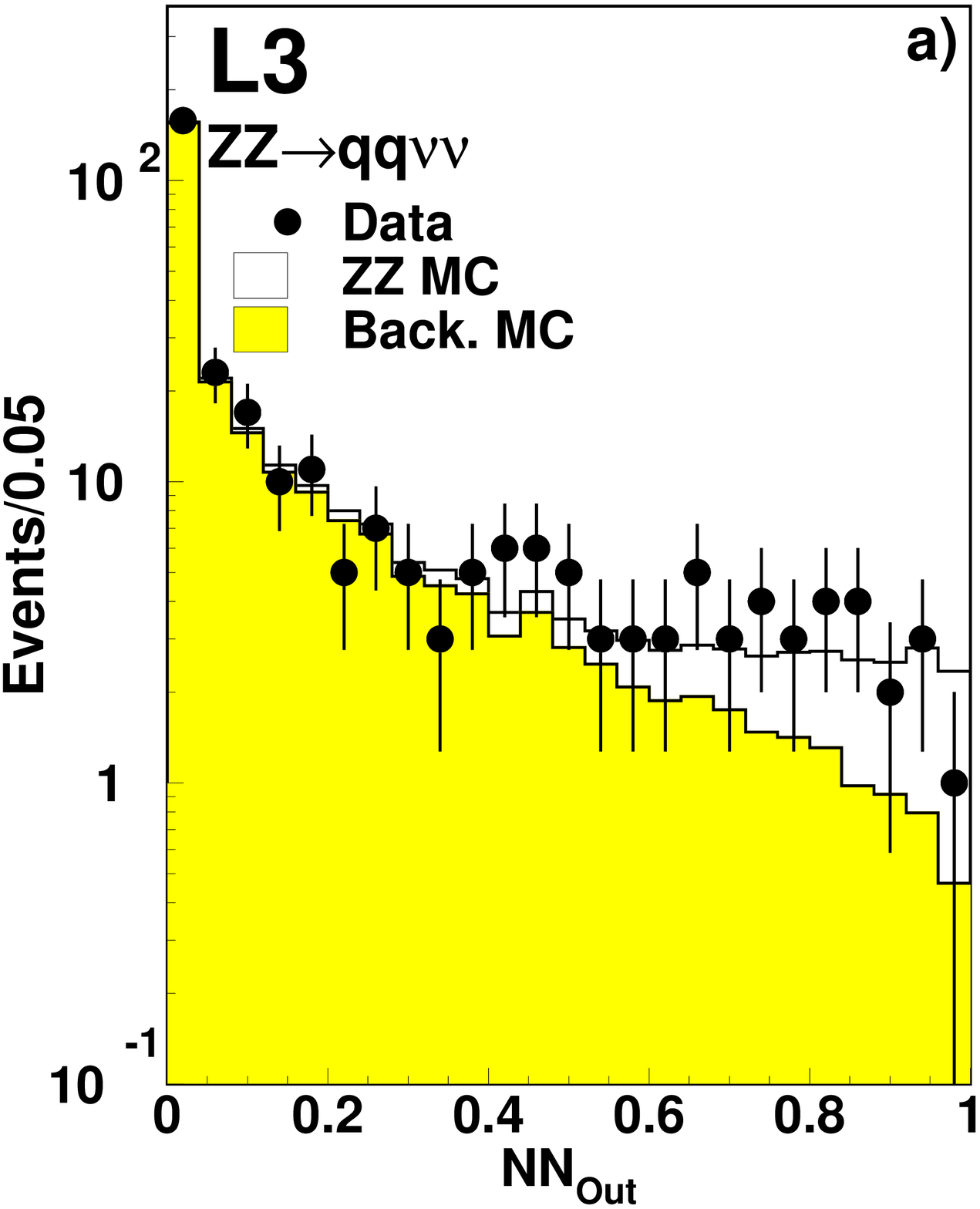}} &
\mbox{\includegraphics[width=8.5truecm]{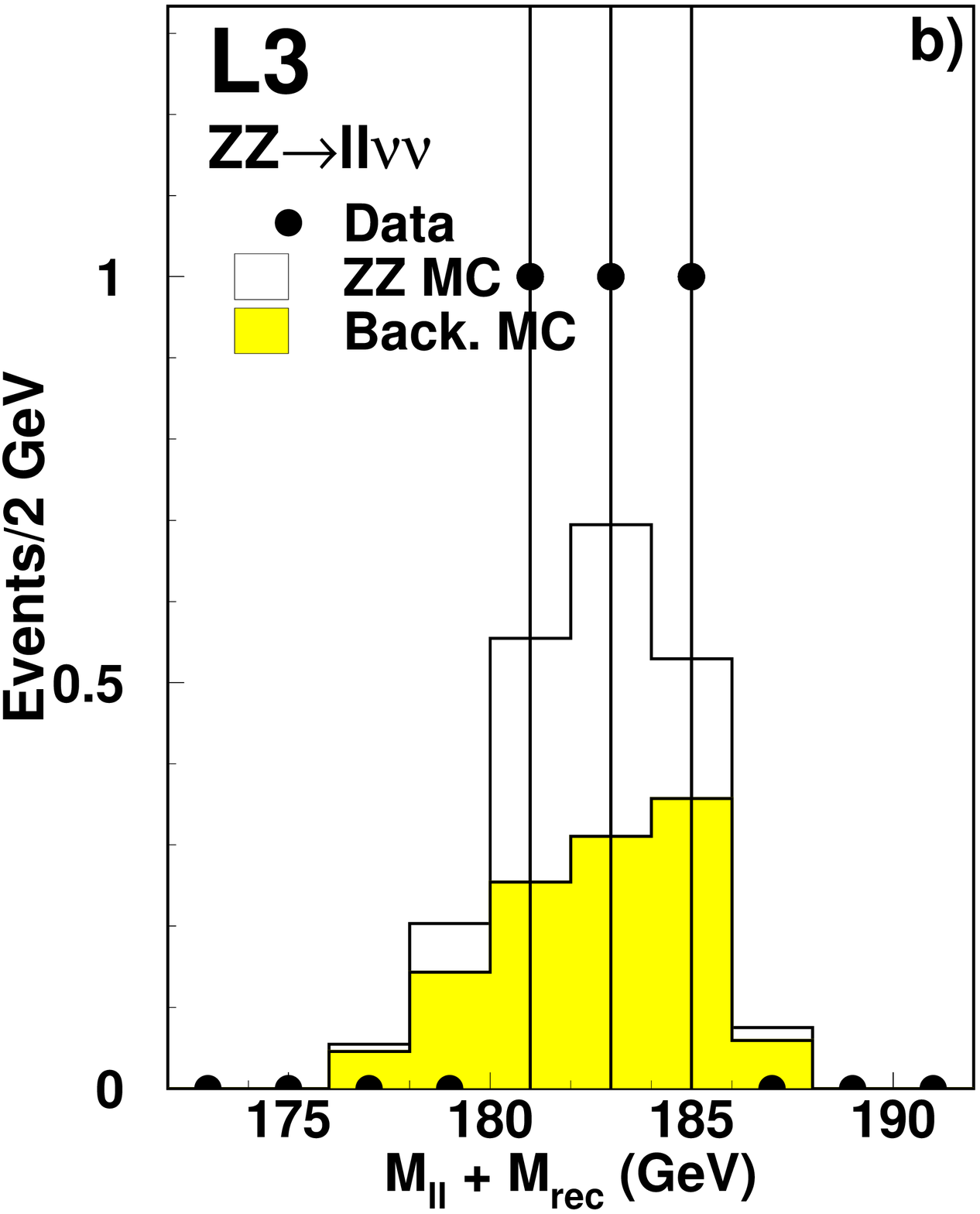}} \\
\mbox{\includegraphics[width=8.5truecm]{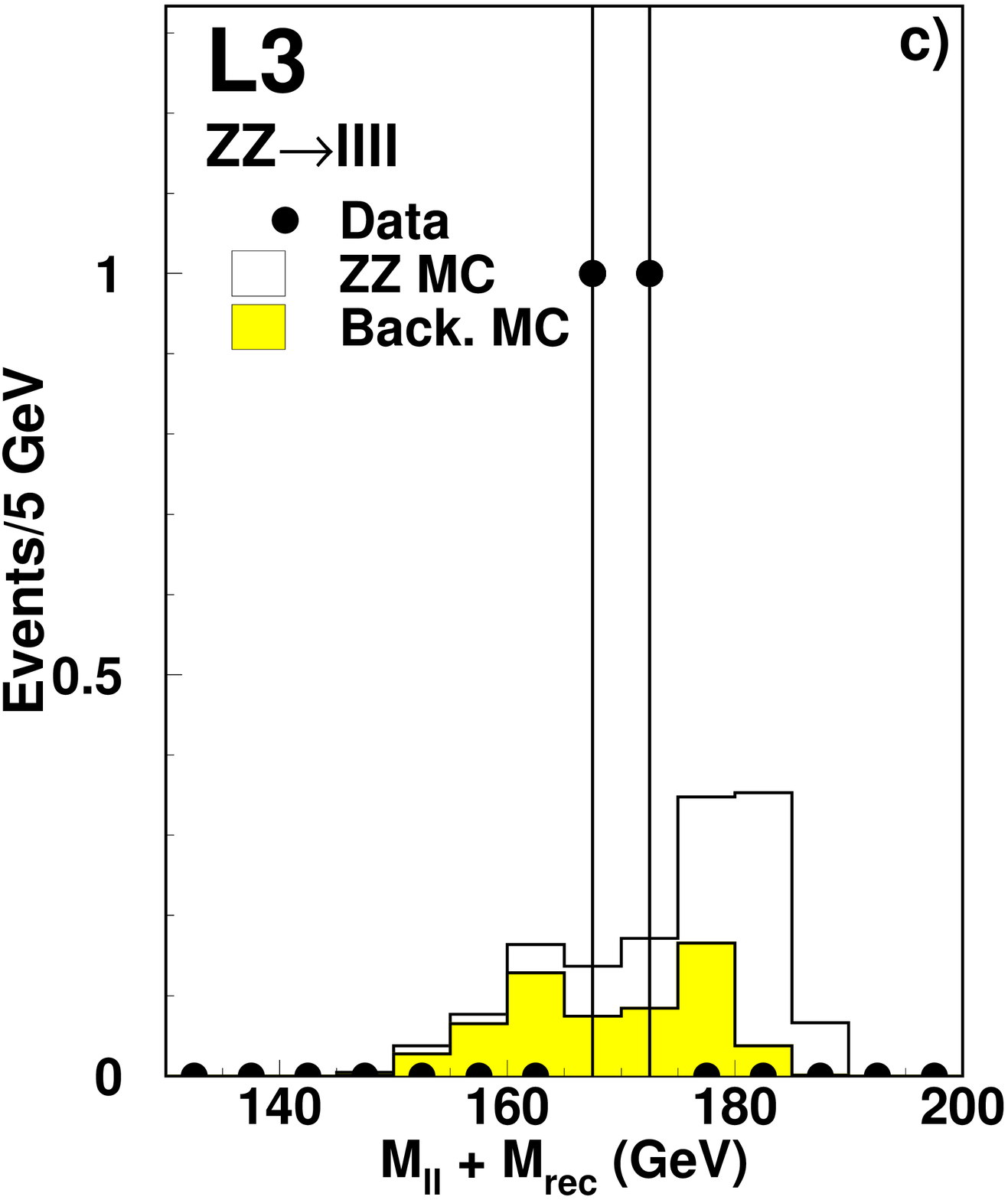}} &
\mbox{\includegraphics[width=8.5truecm]{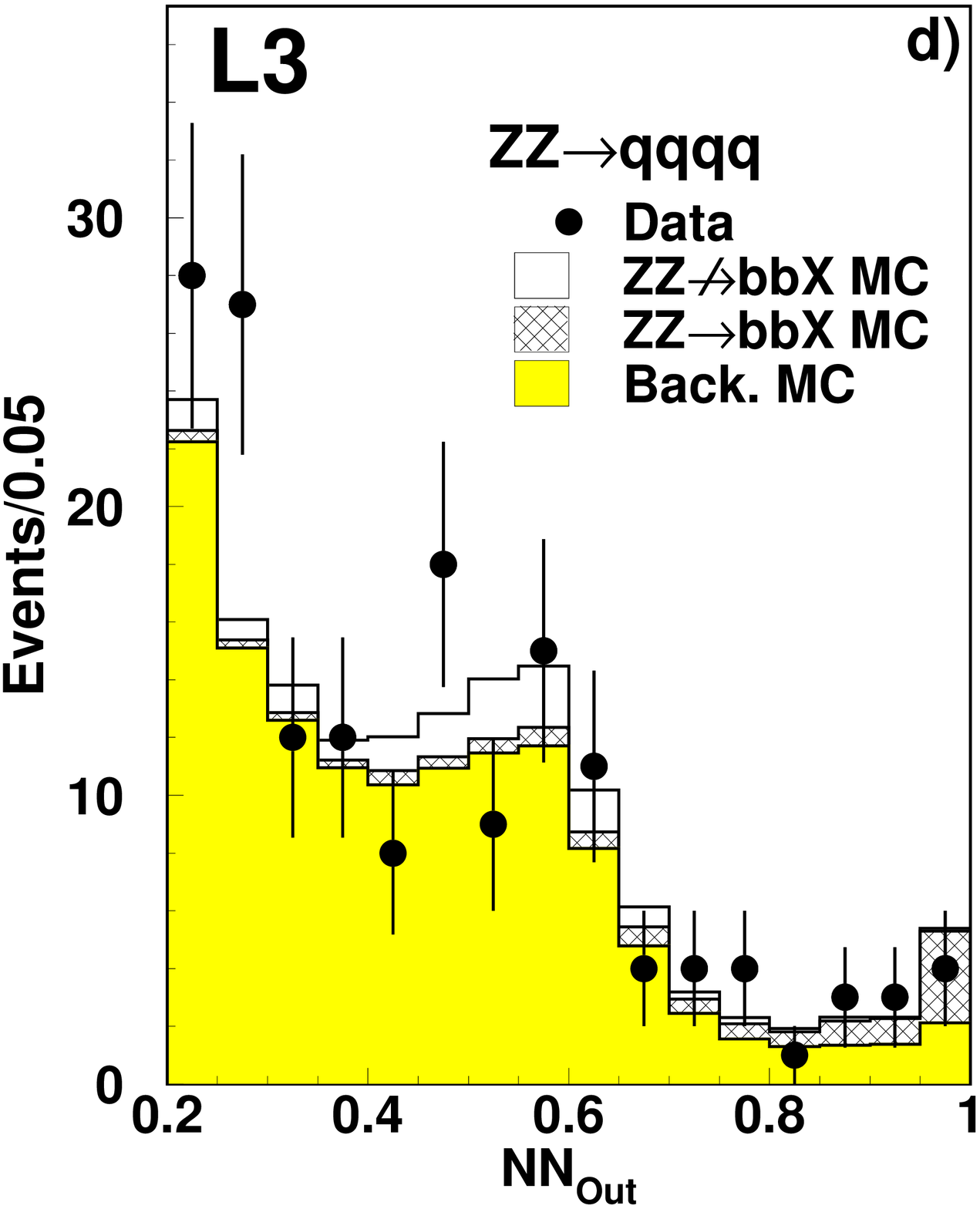}} \\
\end{tabular}
\caption{a) Neural network output for $\qqnn$ selected events. b)
  Sum of the visible and recoil masses for the events selected by the 
  $\llnn$ selection. c) Sum of the invariant and recoil masses of the
  lepton pair  
  closest to M$_{\Zo}$ for the $\llll$ selected events. d)
  Output of the second neural network  for the $\qqqq$ selection;
  signal expectations for events with no or at least one b quark pair
  are presented separately.}
\label{fig:2}
\end{center}
\end{figure}

\newpage
\begin{figure}[p]
\begin{center}
\begin{tabular}{cc}
\mbox{\includegraphics[width=8.5truecm]{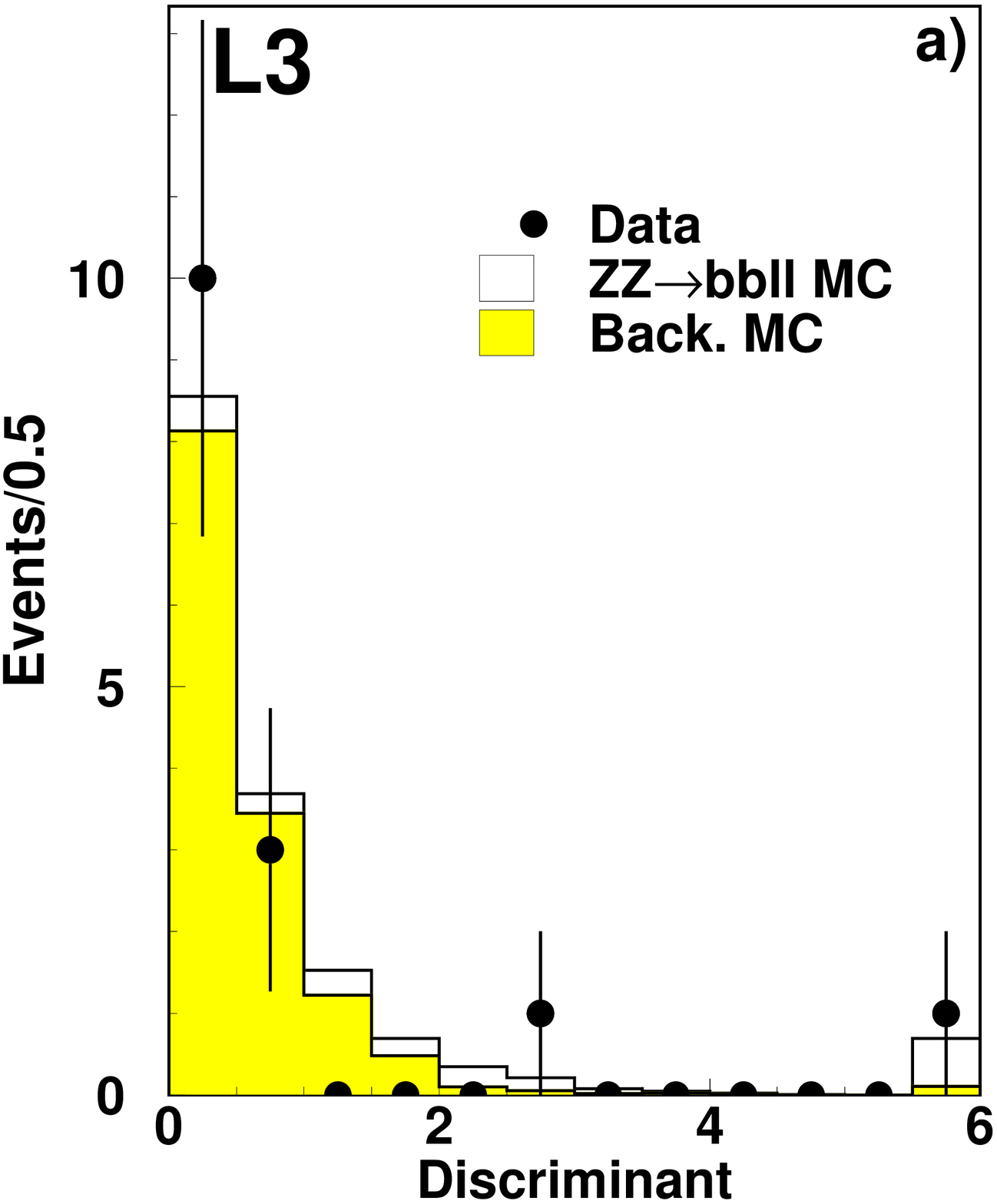}} &
\mbox{\includegraphics[width=8.5truecm]{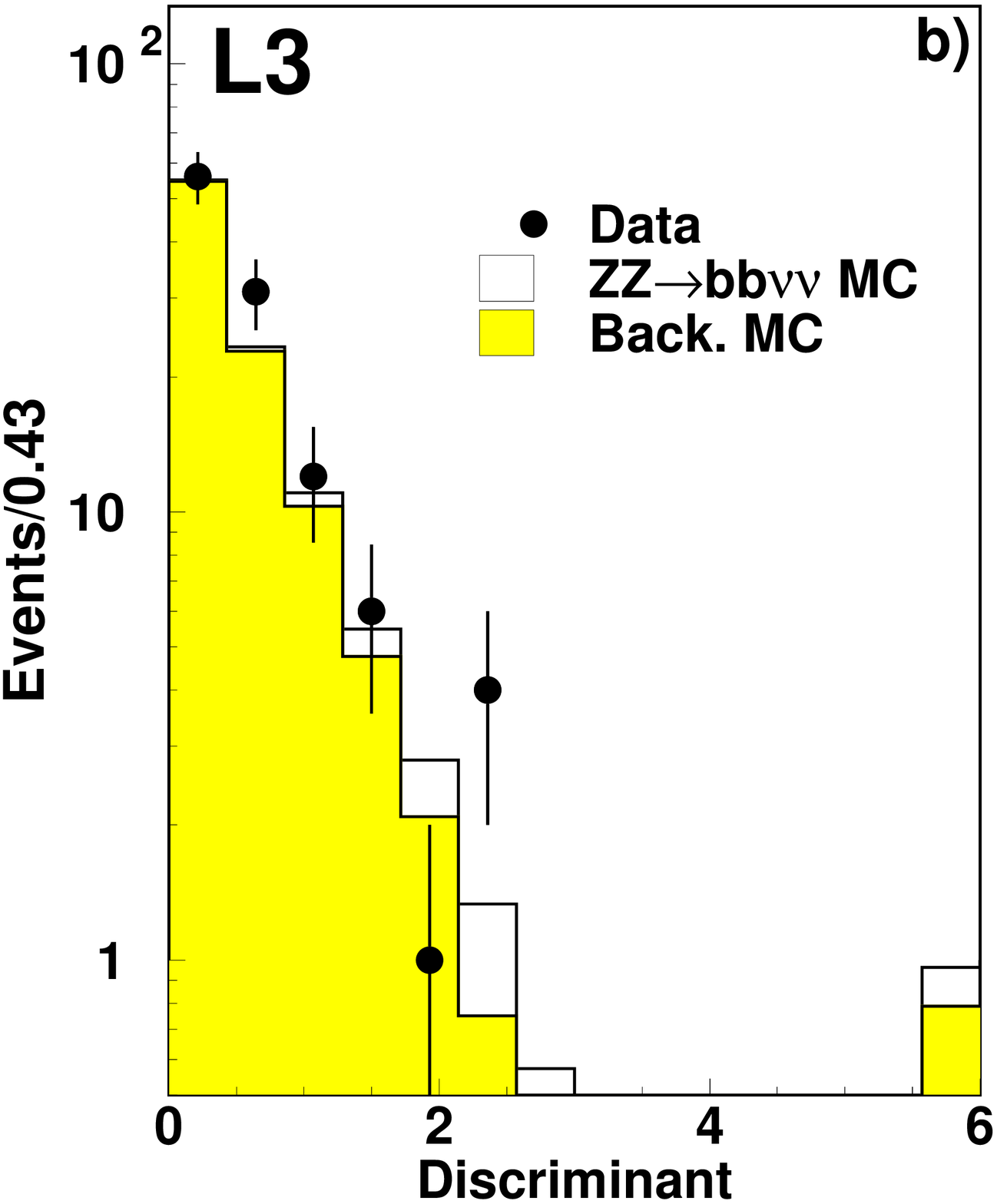}} \\
\end{tabular}
\caption{Discriminant variables for a) the  $\rm b\bar{b}\ell^+\ell^-$
  and b) the $\rm b\bar{b}\nu\bar{\nu}$  selections. The last bin
  shows the overflows.}
\label{fig:3}
\end{center}
\end{figure}

\newpage
\begin{figure}[p]
\begin{center}
\includegraphics[width=15.0truecm]{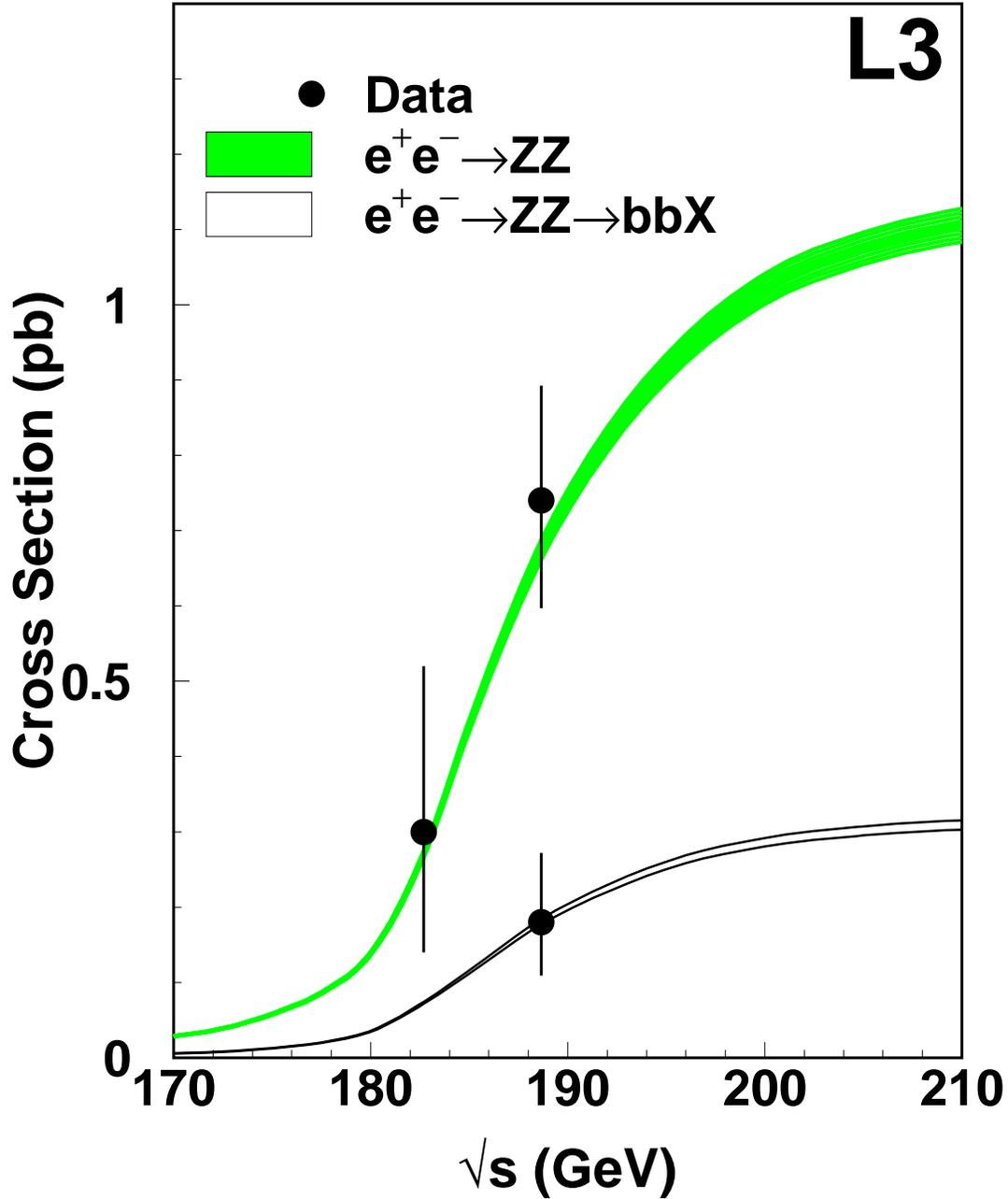}
\caption{Standard Model prediction for the ZZ and $\rm ZZ\ra
  b\bar{b}X$ cross sections and the corresponding 
measurements where statistical and systematic uncertainties are added in
quadrature. Signal definition cuts implemented with the EXCALIBUR
Monte Carlo are applied and a 2\% uncertainty is associated to the
predictions.} 
\label{fig:4}
\end{center}
\end{figure}

\end{document}